\documentclass[conference]{IEEEtran}
\IEEEoverridecommandlockouts
\usepackage{cite}
\usepackage[numbers]{natbib}

\usepackage{amsmath,amssymb,amsfonts}
\usepackage{algorithmic}
\usepackage{graphicx}
\usepackage{textcomp}
\usepackage{xcolor}
\def\BibTeX{{\rm B\kern-.05em{\sc i\kern-.025em b}\kern-.08em
    T\kern-.1667em\lower.7ex\hbox{E}\kern-.125emX}}
\begin{document}

\title{LLM4CodeRE: Generative AI for Code Decompilation Analysis and Reverse Engineering\\
{\footnotesize \textsuperscript{*}LLM4CodeRE: GenAI for RS}
\thanks{Identify applicable funding agency here. If none, delete this.}
}

\author{
\IEEEauthorblockN{
Hamed Jelodar,
Samita Bai,
Tochukwu Emmanuel Nwankwo,
Parisa Hamedi,\\
Mohammad Meymani,
Roozbeh Razavi-Far,
Ali A. Ghorbani
}

\IEEEauthorblockA{
Faculty of Computer Science\\
University of New Brunswick, Canada\\
Emails: \{h.jelodar,samita.bai,tochukwu.nwankwo,parisa.hamedi, mohammad.meymani79,roozbeh.razavi-far, ghorbani\}@unb.ca
}
}

\maketitle

\begin{abstract}
Code decompilation analysis is a fundamental yet challenging task in malware reverse engineering, particularly due to the pervasive use of sophisticated obfuscation techniques. Although recent large language models (LLMs) have shown promise in translating low-level representations into high-level source code, most existing approaches rely on generic code pretraining and lack adaptation to malicious software. We propose \textbf{LLM4CodeRE}, a domain-adaptive LLM framework for bidirectional code reverse engineering that supports both assembly-to-source decompilation and source-to-assembly translation within a unified model. To enable effective task adaptation, we introduce two complementary fine-tuning strategies: (i) a Multi-Adapter approach for task-specific syntactic and semantic alignment, and (ii) a Seq2Seq Unified approach using task-conditioned prefixes to enforce end-to-end generation constraints. Experimental results demonstrate that LLM4CodeRE outperforms existing decompilation tools and general-purpose code models, achieving robust bidirectional generalization.
\end{abstract}


\section{Introduction}

Reverse engineering of portable executable (PE) files, including both benign software and malware, remains a fundamental challenge in cybersecurity. The rapid evolution of modern malware—characterized by pervasive obfuscation, packing, and polymorphism—has made automated code decompilation a critical yet largely unresolved problem. Reverse engineers must routinely translate low-level representations, such as assembly code or raw binaries, into high-level, human-readable source code in order to understand program semantics, uncover malicious intent, and develop effective defensive signatures.\\

Despite the strong semantic modeling capabilities of transformer-based NLP models, reverse engineering pipelines still rely largely on traditional static and dynamic analysis techniques, which are labor-intensive, brittle under obfuscation, and difficult to scale. Even state-of-the-art decompilers struggle with stripped symbols, flattened control flow, and dynamic API resolution commonly used by malware~\citep{tiwari2024cfg, vu2025static}, motivating the need for domain-adaptive learning frameworks that directly model low-level code semantics.

Generative AI models, particularly LLMs, have enabled promising advances in program translation and code understanding by treating code as natural language~\citep{feng2020codebert, hui2024qwen, wang2025asma}. However, most LLM-based decompilation methods rely on generic code pretraining and treat malware as out-of-distribution, limiting their ability to capture malware-specific obfuscation and anti-analysis patterns and reducing reliability in cybersecurity settings~\citep{hu2025sok}.

In this work, we argue that effective GenAI-based reverse engineering and decompilation require domain-specific representation learning and parameter-efficient task adaptation explicitly tailored to both benign and malicious software. In fact,  we introduce LLM4CodeRE, a domain-adaptive LLM framework for bidirectional code reverse engineering that supports both assembly-to-source decompilation and source-to-assembly translation within a unified model.\\\\ By pretraining the backbone model on large-scale corpora of disassembled malware binaries and decompiled pseudo-code, we expose the model to obfuscation techniques, anti-debugging logic, and malware-specific API usage patterns that are characteristic of malicious programs. Inspired by recent advances in domain-adaptive pretraining for code models \citep{jelodar2025llm}, LLM4CodeRE extends this paradigm to the malware domain and, to the best of our knowledge, is the first framework to explicitly target malicious binaries for LLM-based decompilation.


Moreover, we evaluate LLM4CodeRE on bidirectional code transformation tasks, including Assembly-to-Source (Asm→Src) and Source-to-Assembly (Src→Asm), using a unified evaluation protocol that measures three complementary dimensions: (i) semantic similarity, (ii) structural edit similarity, and (iii) re-executability of generated code in a sandboxed environment. Unlike prior work that focuses primarily on static similarity metrics \citep{tan2024llm4decompile, she2024wadec}, our evaluation explicitly tests whether generated programs can be compiled and executed correctly, which is essential for real-world malware analysis and reverse engineering workflows.
Our contributions can be summarized as follows:
\begin{itemize}
    \item To the best of our knowledge, the first malware-aware causal language modeling (CLM) pretraining framework for LLM-based decompilation is introduced, enabling domain-specific representation learning from real-world malicious binaries.
    \item The first bidirectional reverse engineering framework is presented that supports both assembly-to-source decompilation and source-to-assembly translation, using a unified model capable of modeling both benign and malicious code behaviors.
    \item Multi-Adapters(MA) and Seq2Seq(S2S) Unified prefixing are proposed as a hybrid strategy for task-specific adaptation in bidirectional malware code reverse engineering.
    \item A unified evaluation framework is constructed to jointly measure semantic similarity, syntactic fidelity, and re-executability of generated code.

    \item A unified evaluation framework is constructed to jointly measure semantic similarity. The model is available at: https://huggingface.co/JeloH/LLM4CodeRE-S2S-V1

\end{itemize}




\section{Related Work}

\subsection{Malware Decompilation and Reverse Engineering}
The analysis and decompilation of malware binaries has long been a central challenge in cybersecurity. Traditional approaches rely on static analysis techniques such as control-flow graph extraction, function call graph reconstruction, and API usage profiling \citep{tiwari2024cfg, vu2025static}. Dynamic analysis frameworks execute suspicious binaries in sandboxed environments to observe runtime behavior and identify malicious patterns. While effective in controlled settings, these methods are often brittle under obfuscation, packing, and anti-analysis techniques, and they require substantial manual effort from expert analysts. As a result, they struggle to scale to the volume and diversity of modern malware \citep{hu2025sok}.

\subsection{LLMs for Code Understanding and Generation}
Recent advances in large language models have substantially improved automated code understanding and generation. Models such as CodeBERT~\citep{feng2020codebert} and Qwen2.5-Coder~\citep{hui2024qwen} perform well on tasks including code summarization, translation, and synthesis, with instruction tuning further enhancing code reasoning~\citep{wang2025asma}. However, most LLMs are pretrained on benign open-source code and are not explicitly adapted to the statistical and semantic characteristics of malicious software~\citep{jelodar2025llm}.

\subsection{LLMs for Decompilation and Low-Level Code Translation}
Recent work applies LLMs to decompilation and low-level code translation, including LLM4Decompile~\citep{tan2024llm4decompile}, WADEC~\citep{she2024wadec}, and ASMA-Tune~\citep{wang2025asma}. While these approaches demonstrate the feasibility of LLM-based decompilation, they rely on generic or task-specific fine-tuning and do not perform domain-adaptive pretraining on malware corpora.

\subsection{Parameter-Efficient Adaptation of LLMs}
To reduce the cost of fine-tuning large models, parameter-efficient adaptation methods such as adapters \citep{houlsby2019adapter}, prefix-tuning \citep{li2021prefix}, and Low-Rank Adaptation (LoRA) \citep{hu2022lora} have been proposed. These techniques enable task adaptation by introducing lightweight modules or low-rank updates while keeping the backbone parameters frozen. Although widely used in natural language processing and code generation, these methods have not been systematically explored in the context of malware decompilation or evaluated for their impact on both semantic fidelity and executable correctness in low-level code translation tasks.

\subsection{Summary and Positioning}
In contrast to prior work, this study integrates malware-aware CLM pretraining, LoRA-based parameter-efficient fine-tuning, and a unified comparison of Multi-Adapter and Seq2Seq Unified strategies within a single framework. Unlike existing LLM-based decompilers, the proposed approach explicitly targets malicious software as a first-class training domain and evaluates performance using a comprehensive protocol that includes semantic similarity, structural fidelity, and re-executability. This positioning distinguishes the framework from prior efforts that focus primarily on benign code, static similarity metrics, or single adaptation paradigms.

\begin{figure*}[t]
\centering
\includegraphics[width=\textwidth]{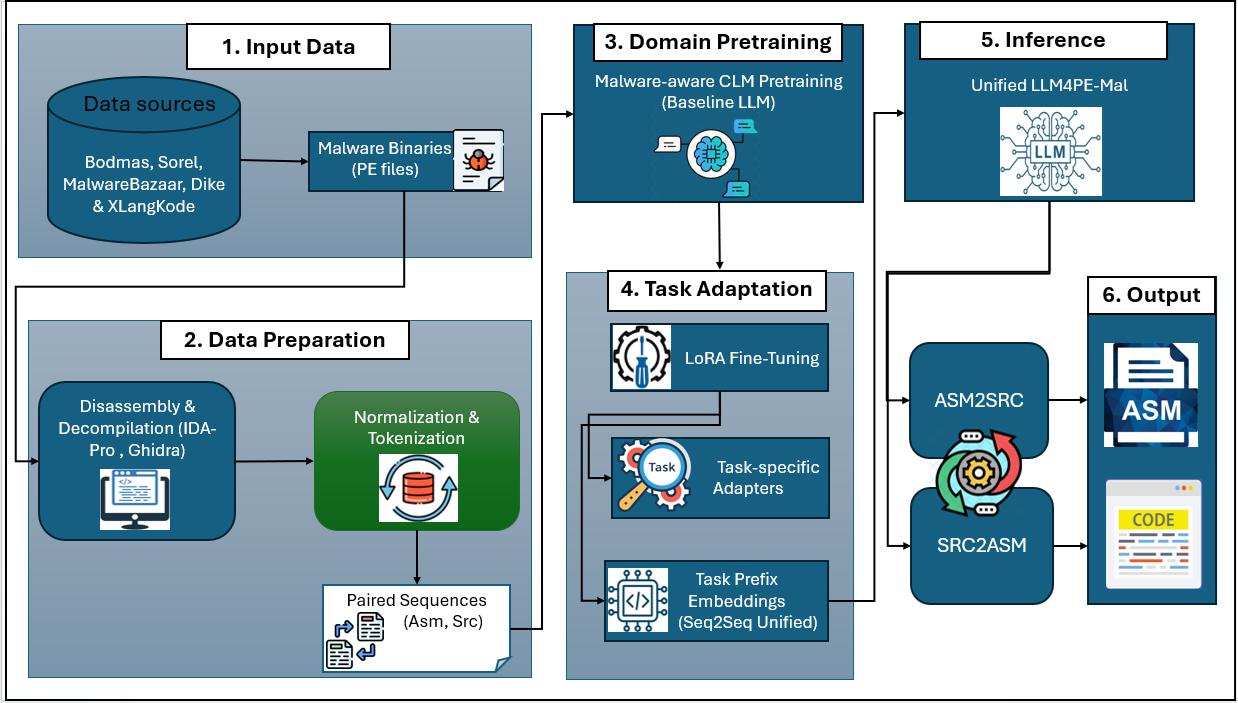}
\caption{System pipeline of the proposed \textbf{LLM4CodeRE} framework. Malware binaries are disassembled and normalized into paired assembly and source representations. A backbone LLM is pretrained using a causal language modeling (CLM) objective on malware corpora. Task adaptation is performed using LoRA and task-specific adapters or Seq2Seq Unified prefix tokens. The unified model supports bidirectional code transformation (Asm$\rightarrow$Src and Src$\rightarrow$Asm) and is evaluated using semantic similarity, edit similarity, and re-executability metrics.}
\label{fig:framework}
\end{figure*}

\section{Methodology}
\label{sec:method}

\subsection{Problem Formulation}
\label{subsec:problem}

Let $\mathcal{D} = \{(x_i, y_i, t_i)\}_{i=1}^{N}$ denote a dataset of paired program representations, where $x_i$ is a low-level input sequence (assembly or binary), $y_i$ is the corresponding high-level output sequence (source code or assembly), and $t_i \in \{\text{Asm}\!\rightarrow\!\text{Src}, \text{Src}\!\rightarrow\!\text{Asm}\}$ denotes the task type.  
The objective is to learn a conditional generation model $P(y \mid x, t)$ that translates between representations under task conditioning.

Given a shared backbone model with parameters $\theta$ and task-specific adaptation parameters $\phi_t$, training minimizes the conditional negative log-likelihood:

\begin{equation}
\mathcal{L}_{\text{task}} = - \sum_{i=1}^{N} \log P(y_i \mid x_i, t_i; \theta, \phi_{t_i}).
\end{equation}

Beyond static similarity, model outputs are evaluated for \emph{functional correctness} by recompiling generated source code and executing it in a sandboxed environment. This formulation explicitly captures both syntactic fidelity and behavioral equivalence, which are essential for real-world malware analysis.

\subsection{Framework Overview}
\label{subsec:overview}

Figure~\ref{fig:framework} illustrates the overall architecture of the proposed \textbf{LLM4CodeRE} framework. The design philosophy decouples domain-specific representation learning from task-level specialization.  
Raw malware binaries are processed using disassembly and decompilation tools (e.g., IDA Pro, Ghidra) to extract assembly code and pseudo-code. These representations are normalized, tokenized, and aligned into paired sequences suitable for causal language modeling and supervised fine-tuning.

The framework consists of three hierarchical adaptation layers:

\begin{enumerate}
  \item A \textbf{malware-aware backbone LLM} obtained via causal language model (CLM) pretraining on real-world malware corpora.
  \item \textbf{Task-specific adapters} that specialize the backbone for different transformation tasks.
  \item \textbf{LoRA low-rank updates} that enable parameter-efficient fine-tuning without overwriting domain knowledge.
\end{enumerate}

Two complementary task adaptation strategies are supported: (i) a \emph{Multi-Adapter} strategy with modular task heads, and (ii) a \emph{Seq2Seq Unified} strategy using task-conditioned prefix tokens.

\subsection{Multi-Adapter Strategy}
\label{subsec:adapters}

The Multi-Adapter strategy introduces lightweight, task-specific parameter modules attached to a shared backbone model. Each adapter is responsible for learning transformations tailored to a specific task, such as Asm$\!\rightarrow$Src or Src$\!\rightarrow$Asm.  
This modular design significantly reduces fine-tuning cost while enabling flexible task specialization.

From an architectural perspective, adapters operate as residual transformation layers. Given a hidden representation $h \in \mathbb{R}^{d}$ from the backbone model, an adapter computes:

\begin{equation}
h' = h + W_2 \, \sigma(W_1 h),
\end{equation}

where $W_1 \in \mathbb{R}^{d \times r}$ and $W_2 \in \mathbb{R}^{r \times d}$ are low-rank projection matrices, $\sigma(\cdot)$ is a non-linear activation function (ReLU), and $r \ll d$ is the adapter bottleneck dimension.

By isolating task-specific knowledge within adapters, the framework avoids catastrophic forgetting when new tasks are introduced. Moreover, adapters can be dynamically activated or combined at inference time, enabling flexible multi-task execution without retraining the backbone.

\subsection{Seq2Seq Unified Strategy}
\label{subsec:seq2seq}

The Seq2Seq Unified strategy relies on a decoder-only causal language model augmented with task-specific prefix tokens. Instead of maintaining separate encoder–decoder pairs for each transformation, all tasks are unified under a single autoregressive decoding framework.

Let $p_t$ denote the learned prefix embedding for task $t$. For an input sequence $x$, the model predicts output tokens autoregressively as:

\begin{equation}
P(y \mid x, t) = \prod_{k=1}^{|y|} P(y_k \mid y_{<k}, x, p_t).
\end{equation}

The prefix conditions the model to perform the desired transformation (e.g., decompilation or recompilation) without modifying the core architecture.   Figure~\ref{fig:performance_pe} shows the relationship between the Multi-Adapter (MA) and Seq2Seq (S2S) strategies, both of which are considered for fine-tuning.

\subsection{Malware-Aware CLM Pretraining}
\label{subsec:clm}

To equip the backbone model with domain-specific knowledge of malicious software, a large-scale causal language model (CLM) pretraining is performed on a curated corpus of real-world malware samples.  
The corpus consists of disassembled assembly code, decompiled pseudo-code, and recovered source-level artifacts extracted from PE binaries collected from public malware repositories and internal threat intelligence feeds.

Each malware sample is normalized via instruction canonicalization, register renaming, and address randomization to reduce syntactic noise. The resulting sequences are tokenized using a hybrid byte-level and instruction-aware tokenizer to preserve both opcode-level semantics and higher-level structural patterns.

Pretraining follows a standard autoregressive objective:

\begin{equation}
\mathcal{L}_{\text{CLM}} = - \sum_{t=1}^{T} \log P(x_t \mid x_{<t}; \theta),
\end{equation}

where $x_t$ denotes the next token in the malware sequence, and $\theta$ represents the backbone model parameters.

\subsection{LoRA-Based Parameter-Efficient Fine-Tuning}
\label{subsec:lora}

To efficiently adapt the malware-pretrained backbone model to downstream code transformation tasks, Low-Rank Adaptation (LoRA)~\citep{hu2022lora} is employed.  
Instead of updating the full parameter matrix $W \in \mathbb{R}^{d \times d}$, LoRA introduces two low-rank matrices $A \in \mathbb{R}^{d \times r}$ and $B \in \mathbb{R}^{r \times d}$ such that:

\begin{equation}
W' = W + \Delta W, \quad \Delta W = B A,
\end{equation}

Where $r \ll d$ is the adaptation rank.

LoRA modules are inserted into the attention and feed-forward projection layers of the decoder backbone. During training, the backbone weights remain frozen, and only the LoRA parameters are optimized.

Compared to full fine-tuning, LoRA significantly reduces memory footprint and training cost while mitigating catastrophic forgetting of malware-domain knowledge acquired during CLM pretraining. In the proposed framework, LoRA complements the Multi-Adapter strategy by capturing fine-grained task variations within each adapter module, yielding a hierarchical adaptation structure consisting of:  
(i) a malware-aware backbone,  
(ii) task-specific adapters, and  
(iii) LoRA low-rank deltas.

\subsection{Training Protocol}
\label{subsec:training}

The training configuration prioritizes efficiency by using a per-device batch size of 64 with 8-step gradient accumulation, yielding a large effective batch size while remaining within GPU memory limits. BF16 precision is employed together with the \texttt{adamw\_torch\_fused} optimizer to reduce computational overhead and improve training stability, enabling efficient optimization with a high LoRA learning rate ($2\times10^{-4}$).


\subsection{Evaluation Metrics}
\label{subsec:metrics}

The framework evaluates decompilation quality using three complementary metrics that capture semantic, syntactic, and functional correctness.

\paragraph{Semantic Similarity}
Semantic fidelity is measured using BERTScore, which computes token-level contextual similarity between generated and reference code sequences.

\paragraph{Edit Similarity}
Structural fidelity is measured using normalized Levenshtein distance:

\begin{equation}
\text{Sim}_{\text{edit}} = 1 - 
\frac{\text{EditDist}(y, \hat{y})}{\max(|y|, |\hat{y}|)}.
\end{equation}

\paragraph{Re-executability.}
Functional correctness is evaluated by recompiling generated source code using GCC tools (v11.3) with optimization level \texttt{-O2}. Successful compilation is followed by sandboxed execution with time and memory limits.  Samples that fail to compile or exceed resource limits are assigned a re-executability score of zero.

\begin{figure*}[t] 
\centering\includegraphics[width=0.45\textwidth]{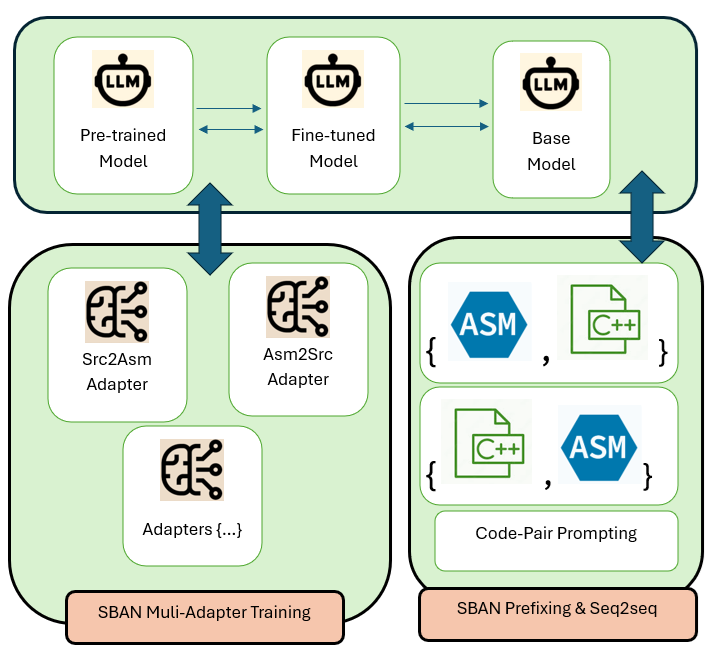} \caption{Relationship between fine-tuning strategies and trained models for decompiling tasks.} \label{fig:fine_tuning} \end{figure*}

\section{Experiments and Results}

\subsection{Experimental Setup}
\label{sec:exp_setup}

\paragraph{Datasets}
We use two complementary datasets for evaluation. For candidate LLM selection and compilation-based comparison, we employ the PE-Machine Learning dataset \cite{pracsec_pe_2021}, consisting of Windows PE malware binaries, which is used solely to compare the compilation count of generated source code across models (Figure~3).

For all downstream experiments—including perplexity, edit similarity, and semantic similarity—we use the SBAN dataset \cite{jelodar2025sban}. SBAN provides aligned malware binaries and decompiler-recovered source representations for Assembly$\rightarrow$Source and Source$\rightarrow$Assembly tasks, aggregating five public malware corpora into 676{,}151 aligned samples (Table~\ref{tab:M-datasets-size}).

\begin{table}[h]
    \centering
     \caption{Statistics of the SBAN Dataset Across Constituent Corpora. \cite{jelodar2025sban}}
    \resizebox{\linewidth}{!}{
    \begin{tabular}{l c c c c}
    \hline
       Dataset  &  Source & NLD & Assembly & Binary\\
       \hline
       1. BODMAS & 93711 & 93711 & 92317 & 88605\\
       2. MalwareBazaar & 14746 & 14746 & 14051 &  13973\\
       3. Sorel20m & 81584 & 81584 &  81177 & 79166\\
       4. Dike & 17431 & 17431 & 12138 & 11726\\
       5. XLangKode & 468679 & 468679 & 5974 &  13299\\
       \hline
       Total &  676151 &  676151 &  205657 & 206769\\
       \hline
    \end{tabular}}
    \label{tab:M-datasets-size}
\end{table}

\paragraph{Baseline Fairness and Comparison Protocol.}
All baselines are evaluated under identical conditions, including tokenization, context length, datasets, and metrics. Backbone selection is performed on a disjoint screening dataset, and no method benefits from additional domain-specific supervision .

\paragraph{Decompilation Pipeline}
All samples are drawn from datasets that were automatically decompiled using Ghidra~11.3. 

\paragraph{Tokenization and Input Length}
  All models use a fixed maximum context length of 1024 tokens. We do not apply any truncation strategy; samples exceeding the token limit are excluded from training and evaluation to preserve full semantic content and prevent partial-function artifacts.

\paragraph{Candidate LLM Models}
We evaluate the following candidate backbone models, as shown in Figure~3:
\begin{itemize}
    \item {DeepSeek-1.3B and DeepSeek-6.7B}
    \item {Qwen-1.5B and Qwen-7B}
    \item {Llama-3.2-1B and Llama-2-7B}
    \item{Mistral-7B}
    \item{Phi-3-small and Phi-4-mini}
\end{itemize}

\begin{figure*}
    \centering
    \includegraphics[width=0.78\textwidth]{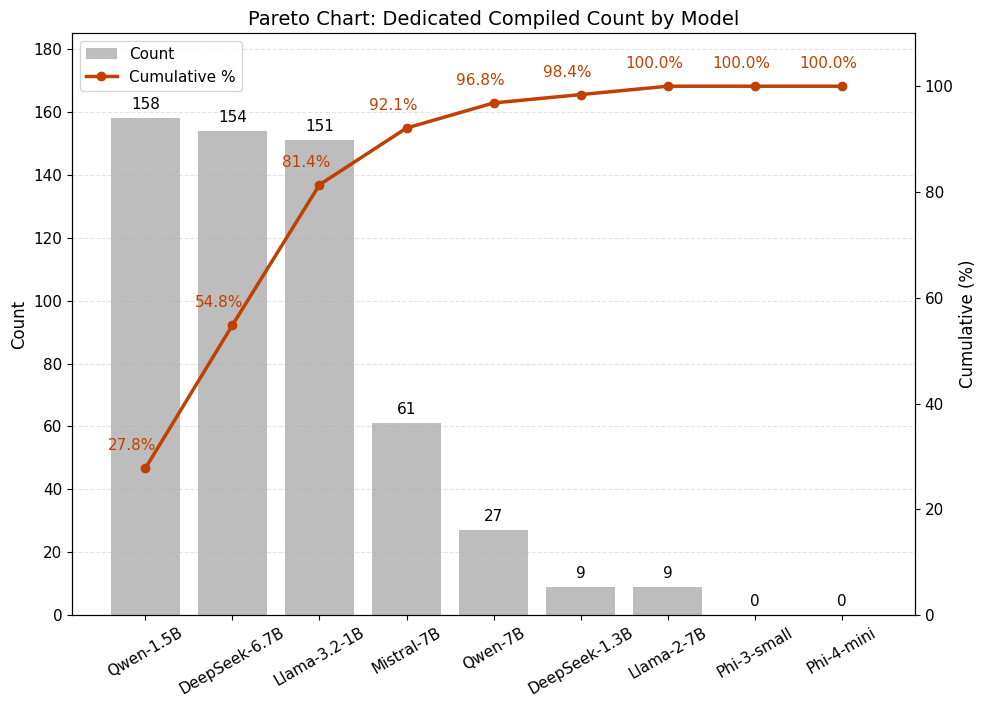}
    \caption{Pareto chart of the dedicated compiled count on the PE-Machine Learning-200 dataset. Bars (left axis) show per-model counts (sorted in descending order), and the line (right axis) shows the cumulative percentage of the total count.}
    \label{fig:performance_pe}
\end{figure*}

For candidate LLM selection, Figure~\ref{fig:performance_pe} shows a Pareto chart of compilation counts on the PE-Machine Learning-200 dataset. The results exhibit a strong concentration effect: the top three models—Qwen-1.5B, DeepSeek-6.7B, and LLaMA-3.2-1B—account for over 80\% of successful compilations, while the top five exceed 96

\begin{figure*}
    \centering
    \includegraphics[width=\linewidth]{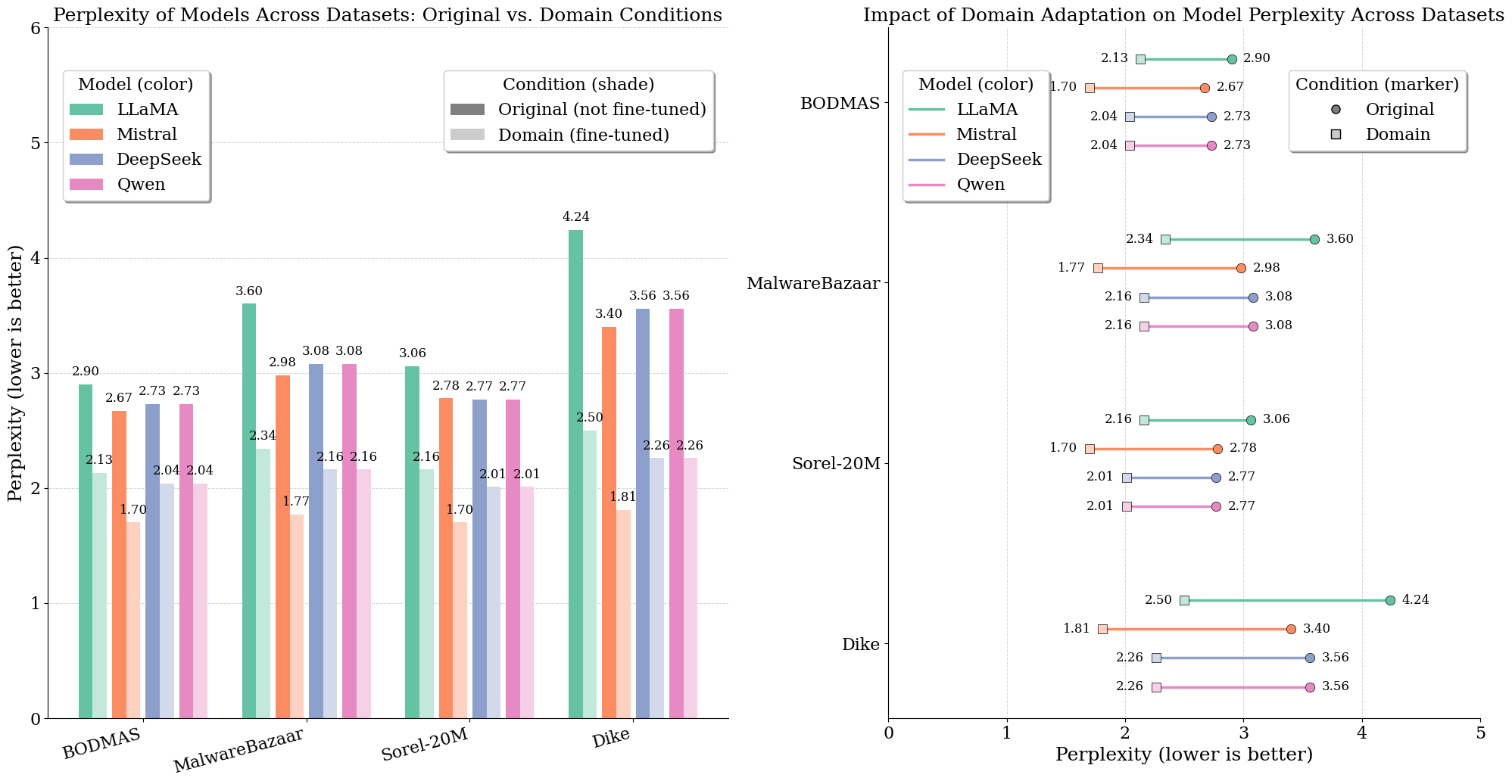}
    \caption{Perplexity comparison across datasets for four language models under two settings: Original (non–fine-tuned) and Domain (domain–fine-tuned). Left: grouped bar chart reporting absolute perplexity on each dataset (lower is better). Right: dumbbell plot illustrating the per-model change from Original to Domain, where shorter (left-shifted) Domain markers indicate improved performance after domain adaptation.}
    \label{fig:perplexity_rate}
\end{figure*}

We further evaluate candidate LLMs using perplexity. Figure~\ref{fig:perplexity_rate} shows that domain adaptation consistently reduces perplexity across all datasets and the four backbone language models. Overall, these results confirm that lightweight domain fine-tuning yields systematic, model-agnostic improvements, motivating its use for subsequent translation and similarity evaluation tasks.

\subsection{Bidirectional Translation Performance}

We assess bidirectional translation performance between assembly and source code using edit similarity and semantic similarity metrics. Figure~\ref{fig:simPlot} compares our \textbf{LLM4CodeRE} models (Multi-Adapter (MA) and Seq2Seq (S2S)) with \textbf{DeepSeek} and \textbf{LLM4Decompile}~\cite{Tan_2024} across both Asm$\rightarrow$Src and Src$\rightarrow$Asm tasks, demonstrating consistent gains from unified bidirectional modeling and domain adaptation.

\subsubsection{Assembly-to-Source (Asm$\rightarrow$Src)}

\begin{figure*}
    \centering
    \includegraphics[width=\linewidth, trim=0cm 0cm 0cm 0cm, clip]{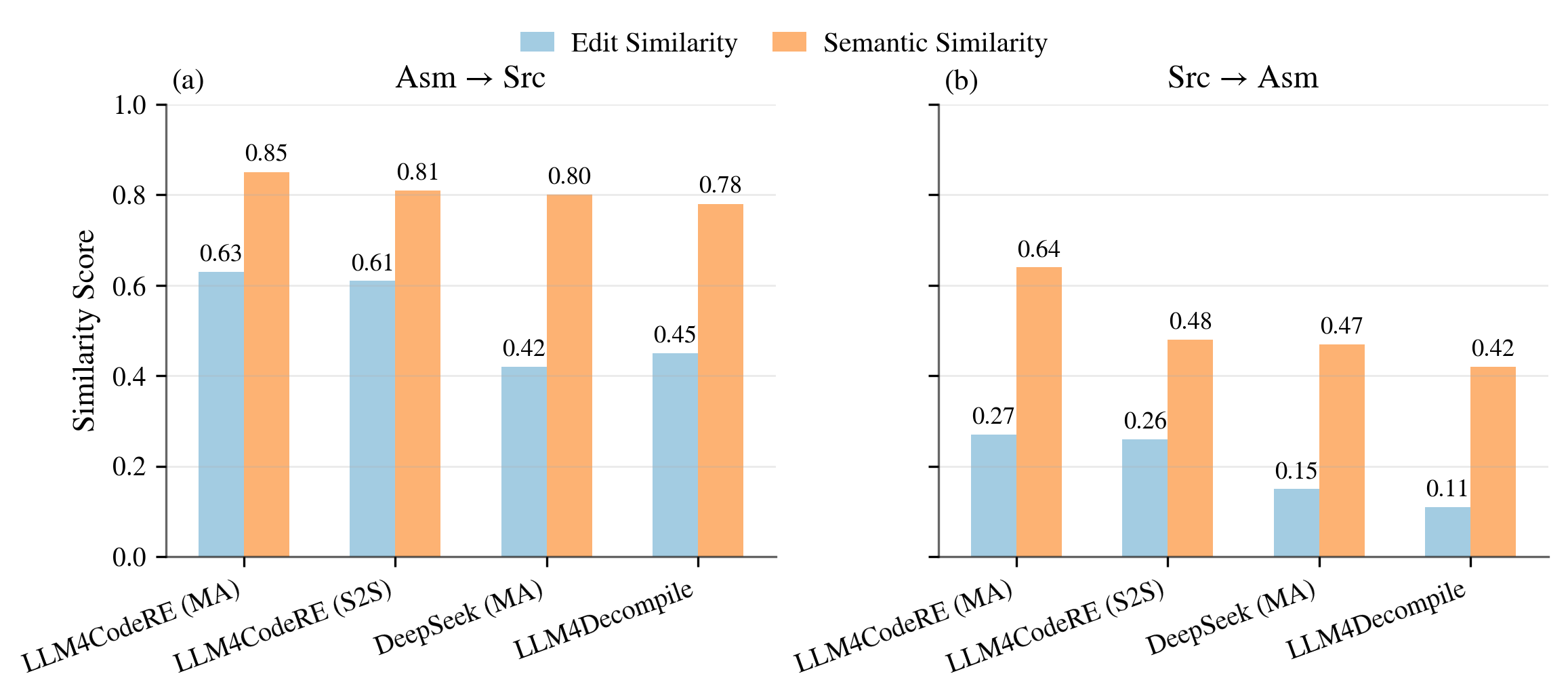}
    \caption{Quantitative evaluation of conversion quality in both directions. (a) Asm-to-Src and (b) Src-to-Asm: Edit similarity and semantic similarity across models (higher is better)}
    \label{fig:simPlot}
\end{figure*}

Figure~\ref{fig:simPlot}(a) shows that both LLM4CodeRE variants—the Multi-Adapter (MA) and Seq2Seq (S2S) strategies—outperform LLM4Decompile and DeepSeek on the Asm$\rightarrow$Src task in terms of semantic and edit similarity. LLM4CodeRE (MA) achieves the highest semantic similarity (0.85) and edit similarity (0.63), followed by the S2S variant (0.81 / 0.61), while DeepSeek and LLM4Decompile obtain lower scores. These results indicate more faithful and semantically consistent source reconstruction with the proposed framework.

\subsubsection{Source-to-Assembly (Src$\rightarrow$Asm)}

Figure~\ref{fig:simPlot}(b) shows that LLM4CodeRE also outperforms DeepSeek and LLM4Decompile on the more challenging Src$\rightarrow$Asm task. The Multi-Adapter (MA) variant achieves the best performance (0.64 semantic, 0.27 edit similarity), followed by the Seq2Seq (S2S) variant, while both baselines obtain substantially lower scores. These results demonstrate the effectiveness of the proposed unified bidirectional training strategy and its robust generalization across translation directions.

\subsubsection{Re-executability Analysis}

\begin{figure*}[t]
    \centering
    \includegraphics[width=0.65\textwidth]{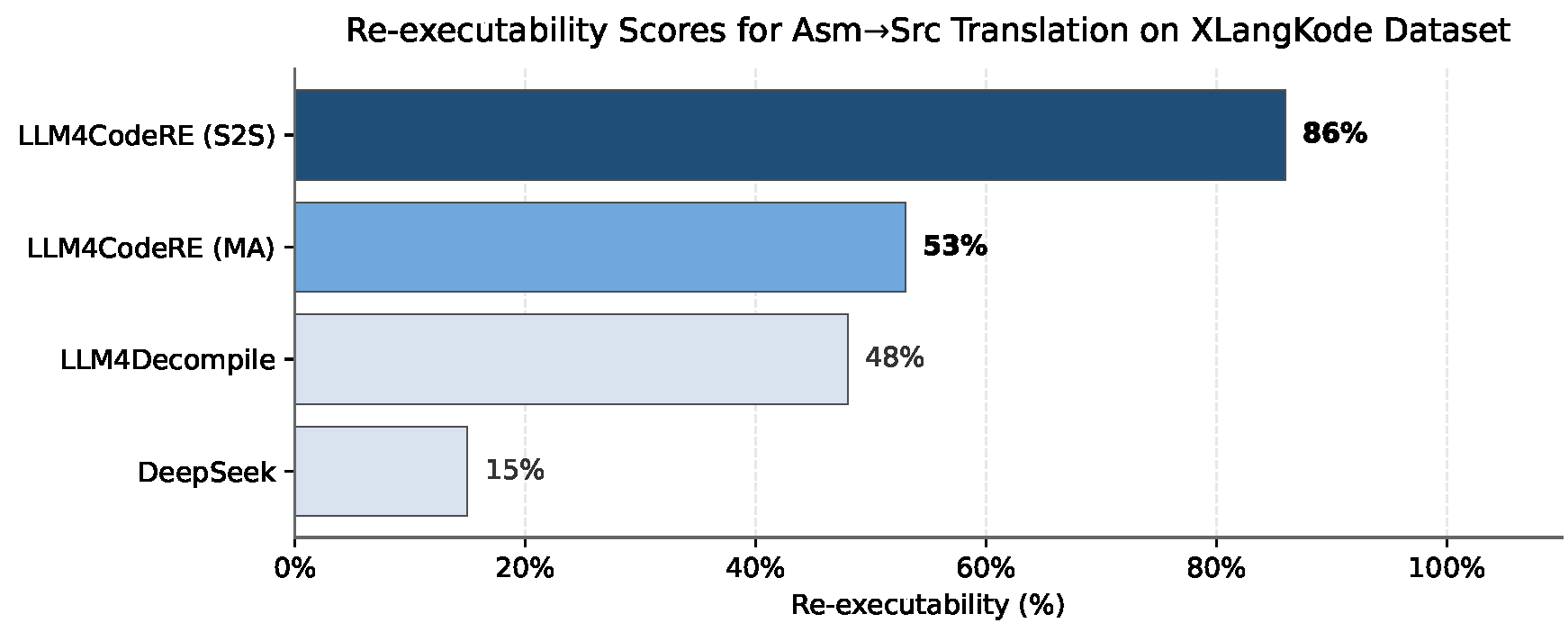}
    \caption{ Re-executability scores (percentage of translated programs that recompile and execute successfully) for assembly-to-source (Asm→Src) translation on the XLangKode dataset. Methods are ranked by performance; LLM4CodeRE (S2S) achieves the highest re-executability (86\%), followed by LLM4CodeRE (MA) (53\%), LLM4Decompile (48\%), and DeepSeek (15\%).}
    \label{fig:reexec_eval}
\end{figure*}

Figure~\ref{fig:reexec_eval} shows re-executability results for the Asm$\rightarrow$Src task on XLangKode. LLM4CodeRE (S2S) achieves the highest re-executability rate (86\%), substantially outperforming LLM4CodeRE (MA) (53\%), LLM4Decompile (48\%), and DeepSeek (15\%). These results highlight the importance of task-aware, unified bidirectional training for generating functionally correct and executable code

\section{Limitations, Ethical Considerations and Future Work}

\subsection{Limitations and Threats to Validity}
The proposed framework primarily targets Windows PE malware and may not generalize to other formats such as ELF or mobile binaries. Automated decompilation introduces label noise, particularly for heavily obfuscated samples. Re-executability is evaluated in a sandboxed environment with limited behavioral coverage, and embedding-based semantic metrics may not fully capture functional equivalence. Future work will explore cross-platform generalization and symbolic execution-based evaluation.

\subsection{Executability Metric Formalization}
Re-executability is defined as successful compilation with GCC (v11.3, \texttt{-O2}) \cite{belinassi2022compiling, shahzad2024neural} followed by sandboxed execution \cite{greamo2011sandboxing} within resource limits. Programs failing compilation or execution receive a zero score, and results are reported as execution success rates.

\subsection{Ground-Truth Approximation and Label Noise}
Source-level references are derived via automated decompilation, providing an approximate \cite{alotaibi2025deep} but standard ground truth. Uniform preprocessing and filtering are applied, and all models are evaluated under identical conditions to ensure fair comparison.

\subsection{Length Bias and Dataset Coverage}
All models use a fixed context length of 1024 tokens, with over-length samples uniformly excluded. The SBAN dataset aggregates multiple heterogeneous malware corpora, promoting robust generalization across diverse malware families.

\subsection{Future Work}
While this work focuses on Windows PE malware, future work will extend LLM4CodeRE to Android malware analysis. This includes supporting Android-specific representations such as APK packages, Dalvik bytecode (DEX), and smali code, as well as modeling Android framework APIs, and permission-based behaviors \cite {li2025foredroid, ibrahim2025lm} . We also plan to evaluate cross-platform generalization between Windows and Android malware in reverse engineering tasks.

\section{Conclusion}

In this work, we introduced LLM4CodeRE, a malware-aware large language model framework for bidirectional code reverse engineering that unifies assembly-to-source decompilation and source-to-assembly translation within a single architecture. 
Extensive experiments on standard datasets demonstrate that LLM4CodeRE consistently improves semantic fidelity, structural alignment, and end-to-end re-executability compared to general-purpose code models and prior decompilation-focused methods.  We believe that LLM4CodeRE provides a strong foundation for future research in executable code generation, automated malware analysis, and robust evaluation of LLM-driven reverse engineering systems.

\bibliographystyle{IEEEtran}
{\small\sloppy
\bibliography{ref}

@inproceedings{feng2020codebert,
    title = "{C}ode{BERT}: A Pre-Trained Model for Programming and Natural Languages",
    author = "Feng, Zhangyin  and
      Guo, Daya  and
      Tang, Duyu  and
      Duan, Nan  and
      Feng, Xiaocheng  and
      Gong, Ming  and
      Shou, Linjun  and
      Qin, Bing  and
      Liu, Ting  and
      Jiang, Daxin  and
      Zhou, Ming",
    editor = "Cohn, Trevor  and
      He, Yulan  and
      Liu, Yang",
    booktitle = "Findings of the Association for Computational Linguistics: EMNLP 2020",
    month = nov,
    year = "2020",
    address = "Online",
    publisher = "Association for Computational Linguistics",
    url = "https://aclanthology.org/2020.findings-emnlp.139/",
    doi = "10.18653/v1/2020.findings-emnlp.139",
    pages = "1536--1547"
}

@inproceedings{li2025foredroid,
  title={ForeDroid: Scenario-Aware Analysis for Android Malware Detection and Explanation},
  author={Li, Jiaming and Chen, Sen and Wu, Chunlian and Zhang, Yuxin and Fan, Lingling},
  booktitle={Proceedings of the 2025 ACM SIGSAC Conference on Computer and Communications Security},
  pages={1379--1393},
  year={2025}
}

@article{ibrahim2025lm,
  title={LM-Scout: Analyzing the Security of Language Model Integration in Android Apps},
  author={Ibrahim, Muhammad and Tuncay, G{\H{u}}liz Seray and Celik, Z Berkay and Machiry, Aravind and Bianchi, Antonio},
  journal={arXiv preprint arXiv:2505.08204},
  year={2025}
}

@inproceedings{alotaibi2025deep,
  title={Deep Learning from Imperfectly Labeled Malware Data},
  author={Alotaibi, Fahad and Goodbrand, Euan and Maffeis, Sergio},
  booktitle={Proceedings of the 2025 ACM SIGSAC Conference on Computer and Communications Security},
  pages={3990--4004},
  year={2025}
}

@article{greamo2011sandboxing,
  title={Sandboxing and virtualization: Modern tools for combating malware},
  author={Greamo, Chris and Ghosh, Anup},
  journal={IEEE Security \& Privacy},
  volume={9},
  number={2},
  pages={79--82},
  year={2011},
  publisher={IEEE}
}

@inproceedings{belinassi2022compiling,
  title={Compiling Files in Parallel: A Study with GCC},
  author={Belinassi, Giuliano and Biener, Richard and Hubi{\v{c}}ka, Jan and Cordeiro, Daniel and Goldman, Alfredo},
  booktitle={2022 International Symposium on Computer Architecture and High Performance Computing Workshops (SBAC-PADW)},
  pages={1--8},
  year={2022},
  organization={IEEE}
}

@inproceedings{shahzad2024neural,
  title={A Neural Network Based GCC Cost Model for Faster Compiler Tuning},
  author={Shahzad, Hafsah and Sanaullah, Ahmed and Arora, Sanjay and Drepper, Ulrich and Herbordt, Martin},
  booktitle={2024 IEEE High Performance Extreme Computing Conference (HPEC)},
  pages={1--9},
  year={2024},
  organization={IEEE}
}

@article{hui2024qwen,
  title={Qwen2.5-Coder Technical Report}, 
      author={Binyuan Hui and Jian Yang and Zeyu Cui and Jiaxi Yang and Dayiheng Liu and Lei Zhang and Tianyu Liu and Jiajun Zhang and Bowen Yu and Keming Lu and Kai Dang and Yang Fan and Yichang Zhang and An Yang and Rui Men and Fei Huang and Bo Zheng and Yibo Miao and Shanghaoran Quan and Yunlong Feng and Xingzhang Ren and Xuancheng Ren and Jingren Zhou and Junyang Lin},
      year={2024},
      eprint={2409.12186},
      archivePrefix={arXiv},
      primaryClass={cs.CL},
      url={https://arxiv.org/abs/2409.12186}, 
}

@article{wang2025asma,
 title={ASMA-Tune: Unlocking LLMs' Assembly Code Comprehension via Structural-Semantic Instruction Tuning}, 
      author={Xinyi Wang and Jiashui Wang and Jinbo Su and Ke Wang and Peng Chen and Yanming Liu and Long Liu and Xiang Li and Yangdong Wang and Qiyuan Chen and Rongze Chen and Chunfu Jia},
      year={2025},
      eprint={2503.11617},
      archivePrefix={arXiv},
      primaryClass={cs.SE},
      url={https://arxiv.org/abs/2503.11617}, 
}

@article{tan2024llm4decompile,
  title={LLM4Decompile: Decompiling Binary Code with Large Language Models},
   url={http://dx.doi.org/10.18653/v1/2024.emnlp-main.203},
   DOI={10.18653/v1/2024.emnlp-main.203},
   booktitle={Proceedings of the 2024 Conference on Empirical Methods in Natural Language Processing},
   publisher={Association for Computational Linguistics},
   author={Tan, Hanzhuo and Luo, Qi and Li, Jing and Zhang, Yuqun},
   year={2024},
   pages={3473–3487} }

@article{she2024wadec,
  title   = {WADEC: Decompiling WebAssembly Using Large Language Models},
  author  = {She, Xiaofei and Zhao, Yao and Wang, Hongyu},
  journal = {Proceedings of the 39th IEEE/ACM International Conference on Automated Software Engineering},
  year    = {2024}
}

@article{hu2025sok,
  title={SoK: Potentials and Challenges of Large Language Models for Reverse Engineering}, 
      author={Xinyu Hu and Zhiwei Fu and Shaocong Xie and Steven H. H. Ding and Philippe Charland},
      year={2025},
      eprint={2509.21821},
      archivePrefix={arXiv},
      primaryClass={cs.CR},
      url={https://arxiv.org/abs/2509.21821}, 
}

@article{jelodar2025llm,
  title={Large Language Models (LLMs) for Source Code Analysis: applications, models and datasets}, 
      author={Hamed Jelodar and Mohammad Meymani and Roozbeh Razavi-Far},
      year={2025},
      eprint={2503.17502},
      archivePrefix={arXiv},
      primaryClass={cs.SE},
      url={https://arxiv.org/abs/2503.17502}, 
}

@inproceedings{houlsby2019adapter,
  title={Parameter-Efficient Transfer Learning for NLP}, 
      author={Neil Houlsby and Andrei Giurgiu and Stanislaw Jastrzebski and Bruna Morrone and Quentin de Laroussilhe and Andrea Gesmundo and Mona Attariyan and Sylvain Gelly},
      year={2019},
      archivePrefix={arXiv},
      url={https://arxiv.org/abs/1902.00751}, 
}

@inproceedings{li2021prefix,
  title={Prefix-Tuning: Optimizing Continuous Prompts for Generation}, 
      author={Xiang Lisa Li and Percy Liang},
      year={2021},
      eprint={2101.00190},
      archivePrefix={arXiv},
      primaryClass={cs.CL},
      url={https://arxiv.org/abs/2101.00190}, 
}

@article{hu2022lora,
  title={LoRA: Low-Rank Adaptation of Large Language Models}, 
      author={Edward J. Hu and Yelong Shen and Phillip Wallis and Zeyuan Allen-Zhu and Yuanzhi Li and Shean Wang and Lu Wang and Weizhu Chen},
      year={2021},
      eprint={2106.09685},
      archivePrefix={arXiv},
      primaryClass={cs.CL},
      url={https://arxiv.org/abs/2106.09685}, 
}

@inproceedings{tiwari2024cfg,
  author={Tiwari, Pradeep Kumar},
  booktitle={2024 2nd International Conference on Device Intelligence, Computing and Communication Technologies (DICCT)}, 
  title={Malware Detection Using Control Flow Graphs}, 
  year={2024},
  volume={},
  number={},
  pages={216-220},
  doi={10.1109/DICCT61038.2024.10532908}}

@article{vu2025static,
   title={A static method for detecting android malware based on directed API call},
  author={Vu Minh, Manh and Nguyen, Huy-Trung and Le, H Viet and Nguyen, Tri Duc and Do, Xuan Cho},
  journal={International Journal of Web Information Systems},
  volume={21},
  number={3},
  pages={183--204},
  year={2025},
  publisher={Emerald Publishing Limited}
}

@misc{jelodar2025sban,
      title={SBAN: A Framework \& Multi-Dimensional Dataset for Large Language Model Pre-Training and Software Code Mining}, 
      author={Hamed Jelodar and Mohammad Meymani and Samita Bai and Roozbeh Razavi-Far and Ali A. Ghorbani},
      year={2025},
      eprint={2510.18936},
      archivePrefix={arXiv},
      primaryClass={cs.IR},
      url={https://arxiv.org/abs/2510.18936}, 
}

@misc{pracsec_pe_2021,
	title = {{PE} {Malware} {Machine} {Learning} {Dataset}},
	url = {https://practicalsecurityanalytics.com/pe-malware-machine-learning-dataset/},
	language = {en-US},
	urldate = {2026-01-22},
	journal = {Practical Security Analytics LLC},
	author = {{pracsec}},
	month = jun,
	year = {2021}
	}

@inproceedings{Tan_2024,
   title={LLM4Decompile: Decompiling Binary Code with Large Language Models},
   url={http://dx.doi.org/10.18653/v1/2024.emnlp-main.203},
   DOI={10.18653/v1/2024.emnlp-main.203},
   booktitle={Proceedings of the 2024 Conference on Empirical Methods in Natural Language Processing},
   publisher={Association for Computational Linguistics},
   author={Tan, Hanzhuo and Luo, Qi and Li, Jing and Zhang, Yuqun},
   year={2024},
   pages={3473–3487} }
}

\end{document}